\documentclass[epj,twocolumn]{webofc}
\usepackage[varg]{txfonts}   
\woctitle{Flavour changing and conserving processes}
\begin{document}
\newcommand{\nc}{\newcommand}
\nc{\be}{\begin{equation}}
\nc{\ee}{\end{equation}}
\nc{\ra}{\rightarrow}
\nc{\gam}{\gamma \gamma}
\nc{\bb}{\bibitem}
\nc{\omg}{\omega}
\nc{\g}{\gamma}
\nc{\pa}{\partial}
\nc{\parsym} {\stackrel{\leftrightarrow}{\pa}}

\title{The HLS approach to $(g-2)_\mu$~: 
A Solution to the ``$\tau$ versus $e^+e^- $'' Puzzle}
%
%

\author{Maurice Benayoun\inst{1}\fnsep
\thanks{\email{benayoun@in2p3.fr}} 
\thanks{\email{maurice.benayoun@gmail.com}} 
}

\institute{LPNHE des Universit\'es Paris VI et Paris VII, IN2P3/CNRS,
4 Place Jussieu, 75252 Paris, France
          }

\abstract{%
 The Hidden Local Symmetry (HLS) Model  provides a framework
able to encompass several physical processes and gives a unfied description of these
in an energy range extending up to the $\phi$ mass.  Supplied with appropriate
symmetry breaking schemes, the HLS Model gives a broken Effective Lagrangian
(BHLS). The BHLS Lagrangian gives rise to a fit procedure in which
a  simultaneous  description of the $e^+ e^-$ annihilations to $\pi^+\pi^-$, $\pi^0
\gamma$, $\eta \gamma$,  $\pi^+\pi^- \pi^0$, $K^+ K^-$, $K_L K_S$ and of the
dipion spectrum in the decay $\tau^\pm \ra \pi^\pm \pi^0 \nu$ can be performed.
 Supplemented with a few pieces 
of information on the $\rho^0-\omg-\phi$ system,
the $\tau$ dipion spectrum  is shown to predict accurately
  the pion form factor in $e^+ e^-$ annihilations.
Physics results derived  from global fits involving or excluding the $\tau$ 
dipion spectra are found consistent with each others. Therefore, 
no obvious mismatch between the $\tau$ and $e^+e^- $
physics properties arises and the $\tau-e^+e^- $ puzzle vanishes
within the broken HLS Model. 
 }
\maketitle
\section{Introduction}
\label{intro}

\indent \indent  
The pion form factor in the $e^+ e^- \ra \pi^+\pi^-$ annihilation ($F_\pi^{ee}(s)$) and 
in the the $\tau^\pm \ra \pi^\pm \pi^0 \nu$ decay ($F_\pi^{\tau}(s)$) are expected to differ 
only by isospin symmetry breaking (IB) terms. Understanding
the relationship between $F_\pi^{ee}(s)$ and $F_\pi^{\tau}(s)$ is important
as it can allow for 2 different evaluations of the dipion contribution to $a_\mu(\pi \pi)$, 
the muon Hadronic Vacuum Polarization (HVP) which could be merged together if  
consistent with each other.  However, this relationship supposes a good 
understanding of isospin symmetry breaking and an appropriate modelling.
 
For a long time \cite{puzzle1,puzzle2}, the comparison between $|F_\pi^{ee}(s)|^2$
and  $|F_\pi^{\tau}(s)|^2$ was not satisfactory and the mismatch \cite{Davier2007}
was severe enough that one started to speak  of a "$e^+ e^-$ vs $\tau$" puzzle.
This has continued up to very recently \cite{DavierHoecker3}.  However, some
works \cite{taupaper,Fred11} indicated that this puzzle could well be a
modelling issue of the isospin symmetry breaking phenomenon. On the other hand,
it was also shown \cite{DavierHoecker} that, numerically, the 
$e^+ e^- - \tau$ discrepancy sensitively depends  on the  $e^+ e^- \ra \pi^+\pi^-$ 
sample considered. Therefore the so--called  $e^+ e^- - \tau$ puzzle may carry
several components. 

The Hidden Local Symmetry (HLS) Model  provides a framework and
a procedure able to address this puzzle in the various aspects just sketched.
The HLS model  encompasses several physical processes and gives a unfied description 
of these in an energy range extending up to the $\phi$ mass.  
However, in order to account precisely for experimental data, it should be supplied with several
symmetry breaking schemes. Among these, an energy dependent mixing mechanism
of the neutral vector meson system ($\rho^0-\omg-\phi$) is generated via loop effects and
allows to define an effective broken HLS (BHLS) model. Within this framework,
 the $e^+ e^-$ annihilations to $\pi^+\pi^-$, $\pi^0
\gamma$, $\eta \gamma$,  $\pi^+\pi^- \pi^0$, $K^+ K^-$, $K_L K_S$ and the
dipion spectrum in the decay $\tau^\pm \ra \pi^\pm \pi^0 \nu$ are
$simultaneously$ accounted for  with the same set of parameters
derived from global fits in procedures involving all the existing data
samples covering the channels listed above.  One can also define a variant
-- named $\tau$ + PDG --
where the $e^+ e^- \ra \pi^+\pi^-$ data are replaced by tabulated $\rho^0$, $\omg$
and $\phi$ particle properties\cite{RPP2012}; with such a tool, one can compare 
the $\tau$ predictions for the $e^+ e^- \ra \pi^+\pi^-$ annihilation and the
data. If the $e^+ e^- - \tau$ puzzle is a relevant concept, this comparison
should enhance the issue.

After a brief review of the BHLS Model and its breaking in Sections \ref{HLS_org},
\ref{breakings}   and \ref{Vmixing}, the various aspects of the global fit method 
are outlined in Section   \ref{global_fit}. Section \ref{taupred} studies the
$\tau$ +PDG predictions and their comparison with the existing $e^+ e^- \ra \pi^+\pi^-$ 
data samples. In Section \ref{globalFits}, one reports on   global fits mixing
$e^+ e^- \ra \pi^+\pi^-$  and $\tau$ dipion data. In Section \ref{hvp}, the effects
of  $\tau$ data on the muon HVP are displayed. Section \ref{conclusion} is devoted to conclusions.

\section{Basics of the  Hidden Local Symmetry Model}
\label{HLS_org}
\indent \indent The Hidden Local Symmetry Model (HLS) Model is a  framework which 
encompasses simultaneously several  different physics processes covered
by a large number of already available data samples. A comprehensive review of 
the  HLS Model is given  in \cite{HLSRef} and a brief account  can be found 
 in \cite{ExtMod3}; however, in order to really  deal with   experimental data at
their present level of accurate, 
breaking procedures need to be implemented.  As these are tightly 
 connected with the HLS Model structure, it  is worth  giving a brief outline of its main
 features.

Beside its non--anomalous sector, which allows to address some
 $e^+ e^-$ annihilation channels and some $\tau$ decays,  
 the HLS Model also contains an anomalous (FKTUY) sector \cite{FKTUY,HLSRef} 
which provides couplings of the form
$VVP$, $VPPP$,  $\gamma PPP$,$VP\gamma$ or  $P\gamma \gamma$
among the light flavor mesons\footnote{In the following, 
$V$ and $P$ denote generically any of the resp. vector and pseudoscalar light flavor 
mesons and, also, the corresponding field matrices without ambiguities.}.
Intrisically, the HLS validity range does not extend much beyond the $\phi$ mass. 

If the $e^+ e^- \ra \pi^+ \pi^-/ K \overline{K}$  annihilations 
or the $\tau^\pm \ra \pi^\pm \pi^0 \nu$
decay clearly proceed from the non--anomalous sector of the HLS model, decays involving,
for instance, $VP\gamma$ or  $P \gamma \gamma$ couplings obviously imply the anomalous
HLS sectors. On the other hand, both the anomalous and non--anomalous sectors of the HLS Model are mandatorily 
requested to account for annihilation channels like
$e^+ e^- \ra \pi^0 \gamma$, $e^+ e^- \ra \eta \gamma$ or 
$e^+ e^- \ra \pi^0 \pi^+\pi^-$. 

The construction of the HLS Lagrangian starts by defining
the (right and left) $\xi$ fields~:
\be
\xi_{R,L} = \displaystyle   \exp{[\pm i \displaystyle P/f_\pi]}
\label{eq1}  
\ee
where $f_\pi$ is the pion decay constant and
$P=P_8+P_0$ is the U(3) matrix of the pseudoscalar fields which includes
the octet and singlet field components \cite{ExtMod3}.
 
The HLS non--anomalous Lagrangian is defined by\footnote{Within the present
one page reminder, we do not discuss the anomalous sectors  and refer the interested
reader to \cite{HLSRef} or \cite{ExtMod3}.}~:
 \be
 \hspace{-0.5cm}
\left \{
 \begin{array}{lll}
 {\cal L}_{HLS}= &{\cal L}_A + a {\cal L}_V & ~~\\[0.3cm]
 {\cal L}_A = &\displaystyle  -\frac{f_\pi^2}{4} {\rm Tr} [(D_\mu\xi_L\xi_L^\dagger -D_\mu\xi_R\xi_R^\dagger)^2] &
 \displaystyle \equiv -\frac{f_\pi^2}{4} {\rm Tr} [L-R]^2\\[0.3cm]
 {\cal L}_V = &\displaystyle  -\frac{f_\pi^2}{4} {\rm Tr} [(D_\mu\xi_L\xi_L^\dagger +D_\mu\xi_R\xi_R^\dagger)^2] &
 \displaystyle \equiv -\frac{f_\pi^2}{4} {\rm Tr} [L+R]^2\\[0.3cm]
 \end{array} 
 \right .
\label{eq2}  
\ee
These expressions involve the covariant derivatives of the $\xi_{R,L} $ fields~:
\be
\left \{
\begin{array}{ccc}
D_\mu \xi_L  = \displaystyle  \pa_\mu \xi_L -i g V_\mu \xi_L +i \xi_L {\cal L}_\mu\\[0.3cm]
D_\mu \xi_R  = \displaystyle  \pa_\mu \xi_R -i g V_\mu \xi_R +i \xi_R {\cal R}_\mu
 \end{array} 
 \right .
\label{eq3}  
\ee
which introduce the usual bare vector field matrix\footnote{which involves the 
so--called ideal combinations $\rho^0_I$, $\omg_I$ and $\phi_I$ 
for the neutral fields.}  $V$;  the other gauge bosons of the Standard Model ($A$, $W^\pm$ 
and  $Z$) are hidden inside ${\cal L}_\mu$ and ${\cal R}_\mu$; neglecting
the influence of the $Z$ boson field absent from the physics we address, these write~:
\be
\left \{
\begin{array}{l}
{\cal L}_\mu =   \displaystyle  e Q A_\mu 
+\frac{g_2}{\sqrt{2}} (W^+_\mu T_+ + W^-_\mu T_-)\\[0.3cm]
{\cal R}_\mu =   \displaystyle e Q A_\mu 
 \end{array}  
 \right .
\label{eq4}  
\ee

The quark charge matrix $Q$ is standard and the matrix $T_+=[T_-]^\dagger$ is constructed out
of matrix elements of the Cabibbo--Kobayashi--Maskawa matrix \cite{HLSRef}. Concerning
the physics parameters,  the above expressions  
exhibit the electric charge $e$, the universal vector coupling $g$
and the weak coupling $g_2$ (related with the Fermi constant by 
$g_2=2 m_W \sqrt{G_F\sqrt{2}}$). Finally, $a$ is a specific HLS parameter
not fixed by the model and expected of the order 2.

\section{Usual symmetry breaking schemes of the HLS Model} 
\label{breakings}
\indent \indent The HLS model obviously provides an elegant unified framework which
covers an important set of annihilation and decay processes. However,
as such,  it cannot  produce a satisfactory account of the real experimental data
falling into its scope. A simple illustration is given by the pion and kaon decay constants
found of equal magnitudes within the unbroken HLS Lagrangian.
 
This clearly indicates that symmetry breaking mechanisms should be supplied. 
The authors of the HLS Model were aware of this difficulty and soon proposed a simple (BKY)
mechanism to break the flavor SU(3) symmetry \cite{BKY} of the model;
this has been   later  extended to include
isospin breaking  \cite{Hashimoto}; for practical purpose, we use the BKY mechanism
as reformulated in  \cite{Heath}. Within the non--anomalous HLS Lagrangian pieces, the
BKY mechanism turns out to perform the substitutions~:
\be
\left \{
 \begin{array}{ccc}
  {\cal L}_A :: & \displaystyle   -\frac{f_\pi^2}{4} {\rm Tr} [(L-R)]^2 \Rightarrow
   -\frac{f_\pi^2}{4} {\rm Tr} [(L-R)X_A]^2\\[0.3cm]
 {\cal L}_V ::  & \displaystyle   -\frac{f_\pi^2}{4} {\rm Tr} [(L+R)]^2 \Rightarrow
  \equiv -\frac{f_\pi^2}{4} {\rm Tr} [(L+R)X_V]^2
 \end{array} 
 \right .
\label{eq5}  
\ee
\noindent where $X_A$ and $X_V$ are real diagonal matrices~:
 \be
\left \{
 \begin{array}{ll}
\displaystyle X_A=& {\rm Diag}(q_A,y_A,z_A)\\[0.3cm]
\displaystyle X_V=& {\rm Diag}(q_V,y_V,z_V)~~~~~.
 \end{array} 
 \right .
\label{eq6}	
\ee
which should be derived from data.

The departures of $q_{A/V}$ and $y_{A/V}$ from 1 measure the isospin symmetry breaking,
while $z_{A/V}$ carry the flavor SU(3) symmetry breaking. Together with determinant 
terms \cite{tHooft} which permit to break nonet symmetry in the pseudoscalar sector, this
provides a reliable description of the light meson radiative decays and of
the $e^+e^- \ra P \g$ annihilations.

\section{Vector field mixing, a new symmetry breaking mechanism } 
\label{Vmixing}
\indent \indent The coupling of the neutral vector mesons carrying no open strangeness to a 
pseudoscalar meson pair is given by the following piece of the  ${\cal L}_V$  HLS  
Lagrangian\footnote{The  isospin breaking effects  generated by the $X_A$ and $X_V$ matrices 
have been removed for clarity; they can be found  in \cite{ExtMod3}.}  \cite{taupaper}~:
 \be
 \begin{array}{ll}
~~& \displaystyle \frac{ia g}{2}  \rho^0_I \pi^- \parsym \pi^+ +
\frac{ia g}{4 z_A}(\rho^0_I + \omg_I -\sqrt{2}  z_V \phi_I ) K^- \parsym K^+ \\
~~& +\displaystyle \frac{ia g}{4z_A}
(\rho^0_I - \omg_I +\sqrt{2}  z_V \phi_I ) K^0 \parsym \overline{K}^0
 \end{array} 
\label{eq7}                                                                                           
\ee
\noindent The last two terms obviously give rise to pseudoscalar kaon loops which
modify the vector mass matrix by $s$--dependent terms. Moreover, the kaon
loops  generate transitions among the ideal $\rho_I$, $\omg_I$ and 
$\phi_I$ of the original Lagrangian and, thus,  give $non-diagonal$ entries
inside the vector meson squared mass matrix \cite{taupaper,ExtMod3}. Stated
otherwise, at one--loop order, the ideal $\rho_I$, $\omg_I$ and $\phi_I$ fields
are no longer mass eigenstates  as expected for the physical  
vector meson fields. The one--loop order mass squared matrix  writes~:
\be 
\left \{
 \begin{array}{lll}
\displaystyle 
M^2(s) &=M^2_0(s)+\delta M^2(s)~~~~{\rm with~:} \\ 
\displaystyle 
M^2_0(s)&={\rm Diag}(m^2+\Pi_{\pi \pi}(s),~m^2,~z_V m^2).
 \end{array} 
 \right .
\label{eq8}                                                                                           
\ee
where  $\delta M^2(s)$ is a non--diagonal perturbation matrix  depending on the kaon loops
to the (otherwise) diagonal matrix $M^2_0(s)$.  The entries of   $M^2_0(s)$
depend on $m^2$ --  the squared $\rho$ (or $\omg$) meson mass, as it occurs in the 
original HLS Lagrangian -- and  on $\Pi_{\pi \pi}(s)$, the dipion loop to which only  $\rho_I$
couples. 

As $\delta M^2(s)$ is  small compared to $M^2_0(s)$ within the range of validity of the HLS model
({\it i.e.} up to the $\phi$ mass region), the eigenvalue problem Eq. (\ref{eq8}) can be solved
perturbatively. The relation between the ideal fields ($V_I$) and the physical fields ($V_R$) 
can be written~:
\be
\left (
\begin{array}{lll}
\rho_{I}\\[0.3cm]
\omg_{I}\\[0.3cm]
\phi_{I}
\end{array}
\right ) = 
\left (
\begin{array}{cll}
\displaystyle  1 &-\alpha  &\beta \\[0.3cm]
\displaystyle  \alpha & 1 &\gamma \\[0.3cm]
\displaystyle -\beta &  -\gamma & 1
\end{array}
\right ) \left (
\begin{array}{cll}
\rho^0_R\\[0.3cm] 
\omg_R\\[0.3cm]
\phi_R
\end{array}
\right ) 
\label{eq9}
\ee
where $\alpha$, $\beta$ and $\gamma$ are functions of $s$, the energy flowing through 
the vector meson line. These functions essentially\footnote{Actually, $K^* \overline{K}$
and $K^* \overline{K^*}$ loops are provided by resp. the $VVP$ anomalous Lagrangian and
by the Yang--Mills term; however, up the $\phi$ mass region they are real
and effectively absorbed inside the subtraction polynomials of the kaon loop combinations.} 
depend on the sum and the difference of 
the charged and neutral kaon loops  and are small 
compared to 1 \cite{taupaper,ExtMod3}. Therefore, the vector field mixing mechanism
introduces breaking terms which are $s$--dependent,  $s$ being the running vector meson
mass.

\begin{figure}
\centering
\includegraphics[width=8cm,clip]{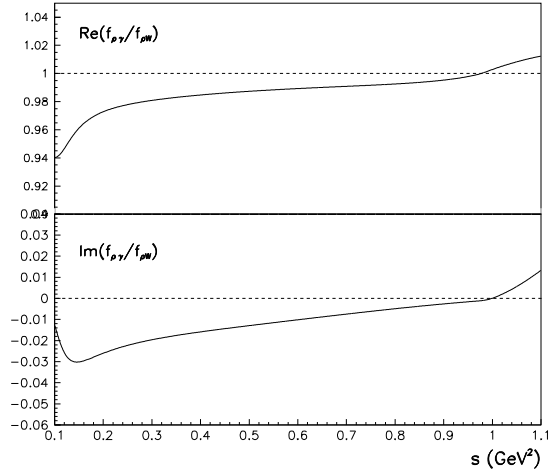}
\caption{ The ratio $f_{\rho^0\g} (s)/f_{\rho^\pm W}$
derived from a BHLS global fit over the region from the two--pion threshold to slightly 
above the $\phi$ mass}.
\label{gamW}      
\end{figure}

This change of fields is what mostly generates an isospin 1 component inside
the physical $ \omg$ and $\phi$ mesons and, then, their couplings to a pion pair as,
using Eq. (\ref{eq9}), 
one gets at first order in breaking parameters\footnote{The $\Delta_V$ comes from an additional (minor) 
breaking process \cite{ExtMod3}; $\Delta_V$  and $\Sigma_V$ are 
combinations of the breaking parameters $q_V$ and $y_V$ generated by the BKY mechanism.}~:
\be
\hspace{-0.8cm}
\begin{array}{ll}
~~&\displaystyle 
 \frac{iag}{2} \rho_I ~\cdot~ \pi^- \parsym \pi^+ \Rightarrow \\
~~&\displaystyle  
 \frac{iag (1+\Sigma_V)}{2}
 \left [
 \rho^0_R +[(1-h_V) \Delta_V - \alpha(s) \omg + \beta(s) \phi \right ] 
 \cdot~ \pi^- \parsym \pi^+
 \end{array}
\label{eq10}
\ee
At this order, the $\rho^0_R$ coupling  to a pion pair is unchanged and remains
identical to those of the $\rho^\pm$ meson. On the other hand, 
the change of fields Eq. (\ref{eq9}) modifies
the Lagrangian coupling  of the neutral $\rho$ meson to the photon, while leaving
unchanged the charged $\rho$ coupling to the $W$ boson. Therefore, the vector field
mixing  makes the ratio of these two couplings $s$--dependent~:                                                                
 \be
\displaystyle 
 \frac{f_{\rho^0\g}(s) }{f_{\rho^\pm W }} = \left [
 1+ \frac{h_V \Delta_V} {3} + \frac{\alpha(s)} {3} + \frac{\sqrt{2} z_V} {3} \beta(s)
 \right ] ~~.
 \label{eq11}
\ee
In order to substantiate this specific breaking of the HLS model, let us quote a result
derived from a global fit involving all the channels listed in the Introduction; the
ratio $f_{\rho^0\g(s)}/f_{\rho^\pm W}$  shown in Figure \ref{gamW} exhibits significant 
variations over the HLS energy range of interest. It is the main mechanism which allows to 
reconcile the $\tau$ dipion spectrum and the $e^+ e^- \ra \pi^+\pi^- $ annihilation cross 
section. These couplings are supplemented by loop corrections \cite{taupaper,ExtMod3}
which also play an important role in defining the effective $\g V$ mixings\footnote{see also 
\cite{Fred11}.}.

On the other hand, the $\tau$ decay process also undergoes several specific breaking 
effects~: The short range \cite{Marciano} and long range \cite{Cirigliano} 
corrections are included when fitting the $\tau$ dipion spectrum, as for the specific 
$\pi^\pm \pi^0$  phase space factor.  These $\tau$ breaking effects  are accounted for 
within fits as usually done;  they are clearly independent of the  HLS breaking mechanisms 
and only come supplementing them.
            
\section{The various aspects of the global fit method}
\label{global_fit}
\indent \indent Gradually equipped since  \cite{taupaper} 
with the various breaking procedures briefly outlined above, 
the HLS Model has evolved toward a broken version (BHLS) \cite{ExtMod3} able 
now to cope simultaneously with  several physics processes, namely 
the $e^+ e^-$ annihilations to $\pi^+\pi^-$, $\pi^0
\gamma$, $\eta \gamma$,  $\pi^+\pi^- \pi^0$, $K^+ K^-$, $K_L K_S$, the
dipion spectrum in the decay $\tau^\pm \ra \pi^\pm \pi^0 \nu$ and, additionally, 
some more radiative decays of light flavor mesons. It involves 25 parameters to 
be extracted from data 
which come intricated simultaneously within the various amplitudes. 

Therefore,   BHLS is  a global model and permits a global fit of the processes
just listed.
As each of the model parameters is involved in several   processes, this gives rise 
to $physics$ correlations  among the various processes belonging to the BHLS realm.
This  also propagates to correlating the data samples covering any of the channels 
involved.
One obviously expects herefrom an improvement of the uncertainties as, for each channel, the available
experimental statistics is practically enhanced by the data collected in any of the 
other channels covered by BHLS. Of course, the fit quality is expected to reflect
that these physics constraints are well accepted by the data.

The number of independent data samples covering the various quoted channels  is 
of the order 50; they are listed and discussed in  \cite{ExtMod3}. In this Reference,
a simultaneous  fit of all available $e^+e^-$  annihilation scan data and of
the published $\tau$ dipion spectra is performed. A very good fit quality is reached 
and no noticeable issue is observed. 
As the model is global, it also represents a new tool to examine precisely several issues
\cite{ExtMod4}~: 
 
\medskip
\begin{enumerate}
\item  The discrepancy  between the dipion spectrum in the $\tau$ decay
and in the $e^+ e^-$ annihilation,
\item  The relative compatibility of the various available $e^+ e^- \ra \pi^+\pi^-$
cross section measurements up to the $\phi$ mass,
\item  The compatibility of the $\tau$ and $e^+ e^-$ based estimates
of  the hadronic vacuum polarization (HVP) contribution to the muon $g-2$. 
\end{enumerate}

The following sections outline the BHLS analysis of these issues.

\section{The BHLS prediction of the pion form factor in $e^+ e^-$ annihilations}
\label{taupred}
\indent \indent Since 2002 several measurements of the pion form factor in 
$e^+ e^-$ annihilations have been published. Beside the data samples collected in scan mode by 
CMD--2 \cite{CMD-2} and SND \cite{SND}, the KLOE Collaboration has produced
three spectra collected in the ISR mode under different conditions, namely KLOE08 \cite{KLOE08},   
KLOE10 \cite{KLOE10} and recently the KLOE12 data sample \cite{KLOE12} -- strongly correlated
with KLOE08. BaBar has also produced a $\pi^+ \pi^-$ spectrum \cite{BaBar} extending up to 1.8 GeV.
Finally, very recently, the BESS III Collaboration has published a new spectrum \cite{BESS-III}
limited to the energy interval $0.6-0.9$ GeV. 
Except for the BESS data sample presented in \cite{ExtMod5}, 
these data samples have been examined  in either of \cite{ExtMod4} or 
\cite{Conf_2013} -- specifically for the KLOE12 sample.
 The present
study  outlines the  treatment of the BESS III sample within the BHLS fitter.

\begin{figure*}
\centering
\includegraphics[width=8cm]{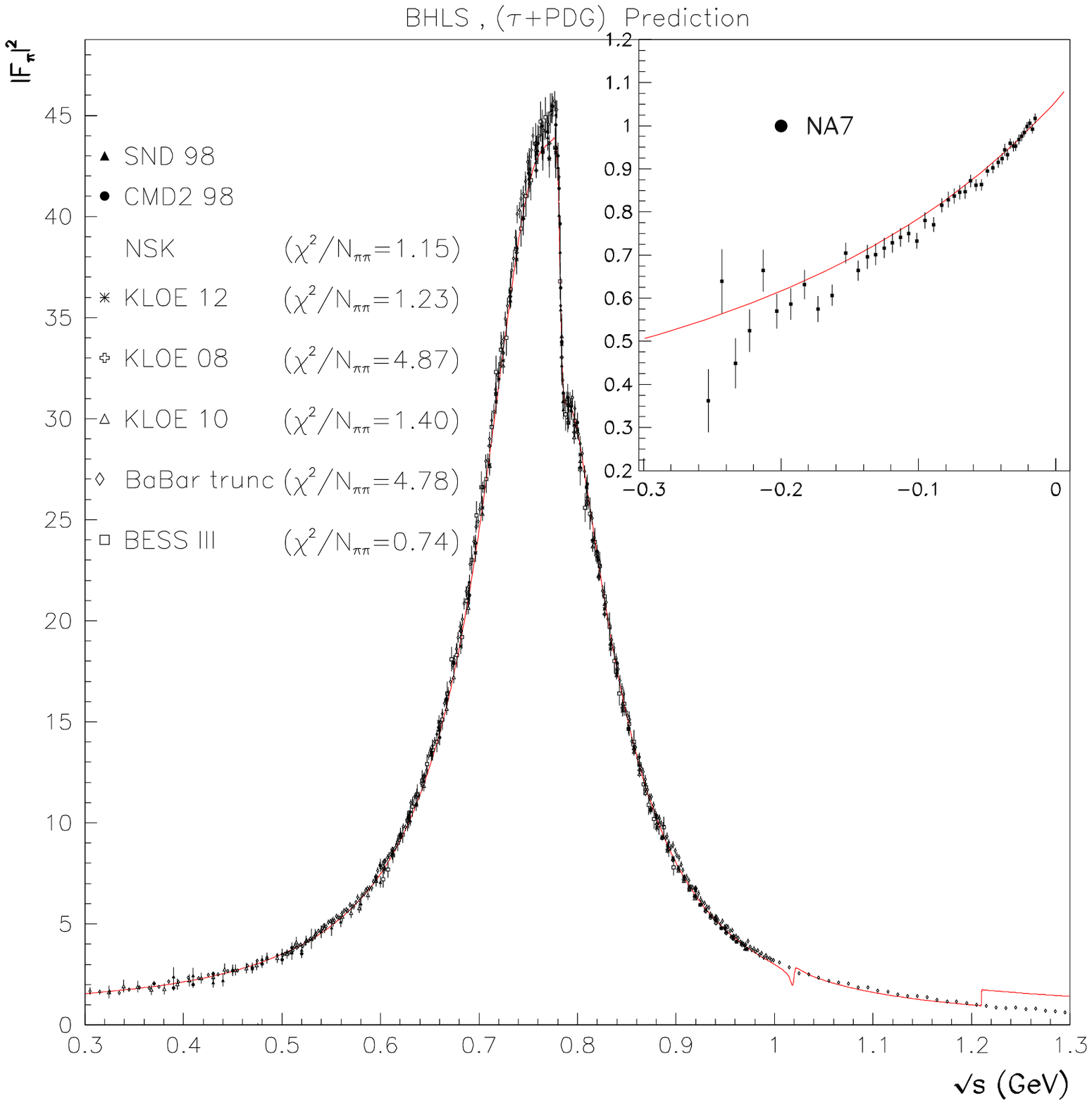}
\hspace{2pc}%
\includegraphics[width=8.cm]{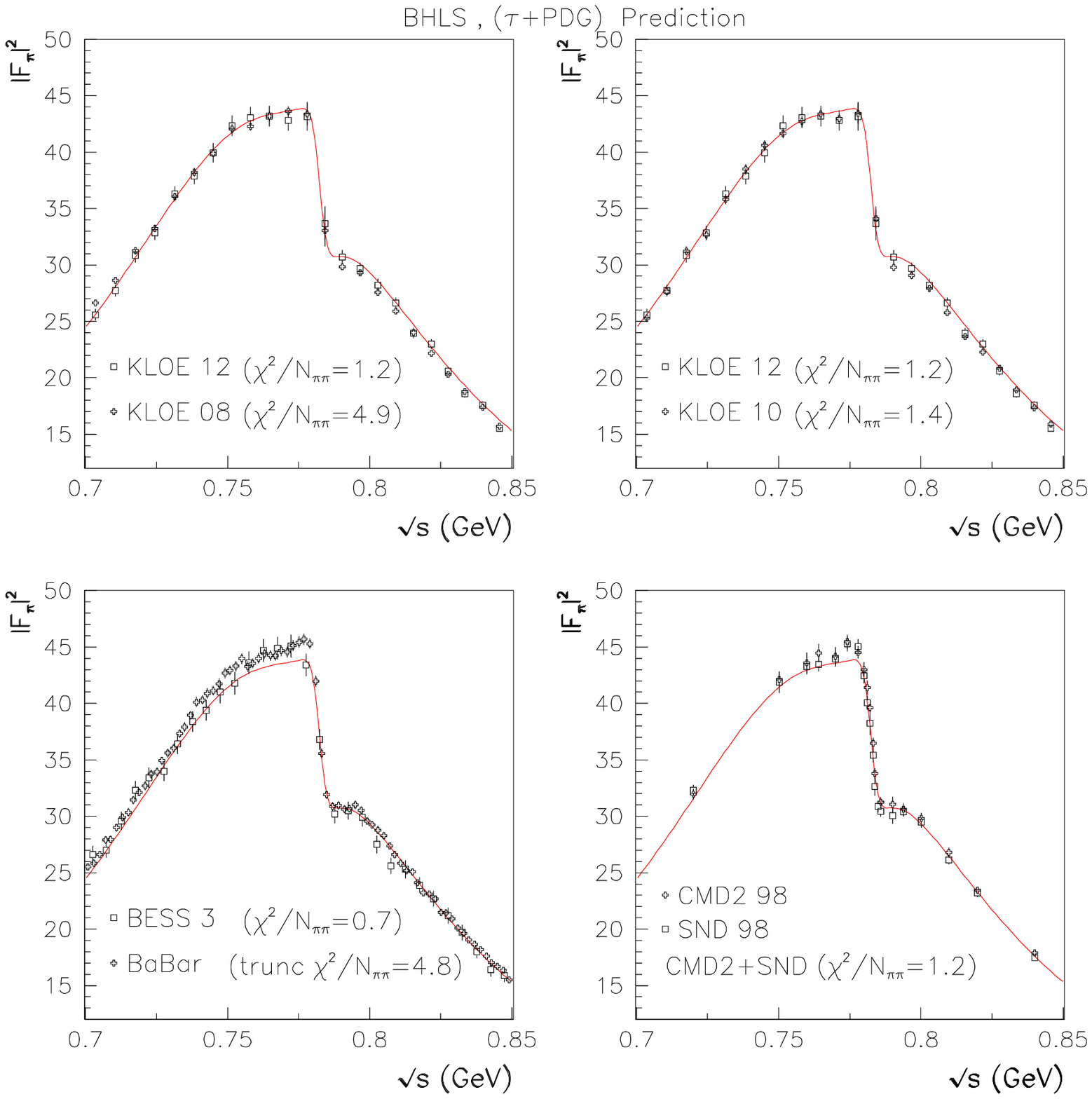}
\caption{Predictions for the pion form factor in the  $e^+e^- \to \pi^+\pi^-$  
 annihilation derived  using the $\tau$ + PDG method. None of the 
 $\pi^+ \pi^-$ spectra is used in the fitting procedure. The leftmost
 pannel displays the overall picture  with data superimposed, the inset showing the form factor
 continuation into the spacelike ($s < 0$)  region. The rightmost pannel displays, 
 enlarged, the behavior of the  various data samples in the $\rho-\omg$ interference region. }
\label{tau_pred}      
\end{figure*}

{\it A priori}, BHLS can  predict the pion form factor in the  $e^+e^- \to \pi^+\pi^-$  
 annihilation relying on the measured $\tau$ dipion spectra \cite{Aleph,Cleo,Belle}, provided
 it is also fed with the appropriate isospin breaking (IB)  information. 
 However, despite the intricacy phenomenon noted above, 
 some of the specific IB  effects  occuring in  the  $e^+e^- \to \pi^+\pi^-$  
 annihilation are marginally constrained by the other annihilation channels
 included within the BHLS realm.  So,  specific data directly reflecting these IB effects
 should be provided to the BHLS fitter.  Such pieces of information are obviously
 related with  the $\omg/\phi \to \pi^+\pi^-$ couplings.  Reference \cite{ExtMod4} proposed to use
 the corresponding tabulated \cite{RPP2012} partial widths and  the (Orsay) phases between the 
 $\omg/\phi$  amplitudes and the underlying coherent $\pi^+\pi^-$ background; this phase information 
 can be  replaced by the relevant tabulated products $\Gamma(V \ra  e^+e^-) \times \Gamma(V \ra  \pi^+ \pi^-)$.  
 Additionally, as the $\rho^0 \to e^+e^-$ coupling is marginally constrained 
 by the non--$\pi^+ \pi^-$  annihilation data, one should include
 the tabulated  $\rho^0 \to e^+e^-$ decay width. This method is named
  $\tau$ +PDG for obvious reasons\footnote{It should nevertheless be kept in mind that 
 the annihilation cross--sections to $\pi^0
\gamma$, $\eta \gamma$,  $\pi^+\pi^- \pi^0$, $K^+ K^-$ and
$K_L K_S$ are always fully involved within the global fit. They mainly serve to fit
the $\g V$ transition amplitudes also involved in the  $e^+e^- \to \pi^+\pi^-$ annihilation process.
}.

For reasons which will become clear soon, it deserves  noting that none of 
the 5 pieces of information supplementing the $\tau$ spectra  in the $\tau$+PDG approach
is influenced  by any of the KLOE, BaBar or BESS III 
$\pi^+\pi^-$ spectra. Actually, they are almost  100\% determined by 
the data collected by the CMD--2 and SND Collaborations.

The red curve in Figure \ref{tau_pred} displays the  $\tau$ +PDG distribution and,
superimposed,  all the available $e^+e^- \to \pi^+\pi^-$   data samples. When 
{\sc minuit}
has converged, one can compute the $\chi^2$ distance of each of the available
$\pi^+ \pi^-$ spectra to the $\tau$ +PDG best fit function; the average 
$\chi^2$ per data point is then calculated for each data sample and all are reported inside
the plots as comments\footnote{Quite generally, the CMD--2 and SND data samples are grouped
together and denoted NSK; to avoid energy calibration issues around the $\omg$ mass, 
the   $\chi^2$ value shown for BaBar is calculated by amputating the
 spectrum from its part falling between 0.76 and 0.80 GeV.}.

 Obviously, the
left--hand side pannel indicates that, overall, the agreement between the
 $\tau$ +PDG $prediction$
and the data is satisfactory, and the inset indicates that this agreement 
extends to
the close spacelike region\footnote{It is nevertheless premature to include
this region inside the BHLS fit \cite{ExtMod5}.}. 
Therefore, the accepted (PDG) values for the IB pieces of information  listed above allow to
recover the gross features of the pion form factor with a noticeable precision;
this already indicates that the IB mechanism as plugged within BHLS is appropriate.

The right--hand side pannel, however, indicates that we are faced with a contrasting 
picture, depending on the data samples examined.  As a good tag of the agreement
between the BHLS $\tau$ +PDG and the (secure)  NSK data\footnote{The NSK data are implicitly
(or explicitly) considered as a reference, as they can accomodate $separately$ almost
all the other data samples with, however, various qualities.}, the relevant
subpannel in the right--hand side of Figure \ref{tau_pred} displays the average
$\chi^2$ distance per data point of the NSK samples; one gets
$\overline{\chi}^2_{NSK}=1.2$, close to the best fit value in a fit where the PDG
information is replaced by the CMD--2 and SND  $e^+e^- \to \pi^+\pi^-$ 
data samples \cite{ExtMod3}. This also gives a hint about  the range of acceptable values
for the $\overline{\chi}^2$ associated with any given sample.

Therefore,
the BHLS fit in the $\tau$+PDG mode provides IB parameter values which allow
the underlying (BHLS) IB framework to exhibit a full consistency of all
non $\pi^+\pi^-$ data
with the $\pi^+\pi^-$ NSK ({\it i.e.} CMD--2 \& SND) data. 
The picture is clearly alike for the KLOE10 ($\overline{\chi}^2_{KLOE10}=1.4$), KLOE12
($\overline{\chi}^2_{KLOE12}=1.2$) and BESSIII ($\overline{\chi}^2_{BESS}=0.7$) data samples.
In contrast, one observes that the KLOE08 ($\overline{\chi}^2_{KLOE08}=4.9$) or BaBar
($\overline{\chi}^2_{BaBar~trunc}=4.8$) samples are farther than could be expected. 

The upper left--hand plot in the right pannel of Figure \ref{tau_pred} is also quite
informative; indeed it shows that the twin samples KLOE08 and KLOE12 \cite{KLOE08,KLOE12} carry central 
values very close to each other and  that both follow almost exactly the $\tau$ +PDG predicted curve.
Nevertheless, KLOE12 exhibits a $\overline{\chi}^2$ value  in close agreement  with the $\tau$ +PDG
expectations while  KLOE08 does not. This should be due to the estimates
and structure of the reported correlated systematic uncertainties, seemingly better understood
for the  KLOE12 sample. 

Finally, there is a
clear contradiction between the KLOE and BaBar samples, 
as already reported by other authors  (see for instance \cite{DavierHoecker})
but, also,   BaBar does not fit well with the BHLS $\tau$ +PDG predictions.

Therefore, as in the comparison between the $\tau$ +PDG predictions and
 the various   $\pi^+\pi^-$  data samples, 
five out of the seven available  independent $\pi^+\pi^-$ samples do not exhibit 
any kind of mismatch, one is obviously tempted to conclude that there is  no  evidence for 
a  $\tau-e^+e^-$ puzzle.
The BHLS approach would rather indicate that the reported puzzle comes from  
a non--adequate IB modelling.

\begin{table*}
\caption{Fit mixing the indicated $e^+e^- \to \pi^+\pi^-$  samples
$and$ the $\tau$ spectra in single mode or combined.  The values are the ratios
$\chi^2/N_{\pi^+\pi^-}$ returned by the fits and the global fit probability 
are given for each data sample or for the selected sample combinations. 
} 
\centering
\label{Table:T1} 
\begin{tabular}{llllllll}                             
\hline
Fit Cond. & KLOE08 & KLOE10  & KLOE12 & NSK & BESS  & BaBaR & BaBaR \\ 
{\small $\chi^2/N_{\pi^+\pi^-}$} & ~~ & ~~    & ~~   & ~~ & ~~   &(trunc)&(full)  \\\hline 
Single   &1.64  & 0.96  &1.02  &0.96 & 0.56 & 1.15 & 1.25\\
    (prob) &  (59\%) &   (97\%)  &   (97\%)   &   (97\%) &  (99\%)&  (74\%)&  (40\%)\\\hline 
Comb. 1 (0.98 [99\%]) & $-$ &1.00  & 1.05  & 1.11  & 0.61 &$-$ &$-$\\
Comb. 2 (1.06 [97\%]) & $-$ &1.02  & 1.05  & 1.10  & $-$ &$-$ &$-$\\
Comb. 3 (1.21 [22\%]) & $-$ &1.01  & 1.54  & 1.18  & 0.56 &$1.36$&$-$\\
\hline 
\end{tabular}
\end{table*}

\section{The BHLS global fits}
\label{globalFits}
\indent \indent So,  the comparison between the $\tau$ predictions -- also based
on commonly accepted PDG information -- and the various $\pi^+\pi^-$ data samples
indicates various behaviors. More precisely, the  $\tau$ + PDG method gives a well--founded
indication that the
CMD--2, SND, KLOE10, KLOE12 and BESSIII $e^+e^- \to \pi^+\pi^-$ data samples should be quite consistent
with the $\tau$ dipion spectra; in contrast, one may expect that 
 KLOE08 and BaBar should exhibit some difficulty to accomodate
 the $\tau$ spectra within the BHLS framework. A step further is to perform global fits
 within the BHLS framework including the $\tau$ spectra and the various 
$e^+e^- \to \pi^+\pi^-$ data samples, each in isolation or in combinations. The results obtained
should allow for more conclusive statements.

\begin{figure*}[t]
\centering
\includegraphics[width=8cm]{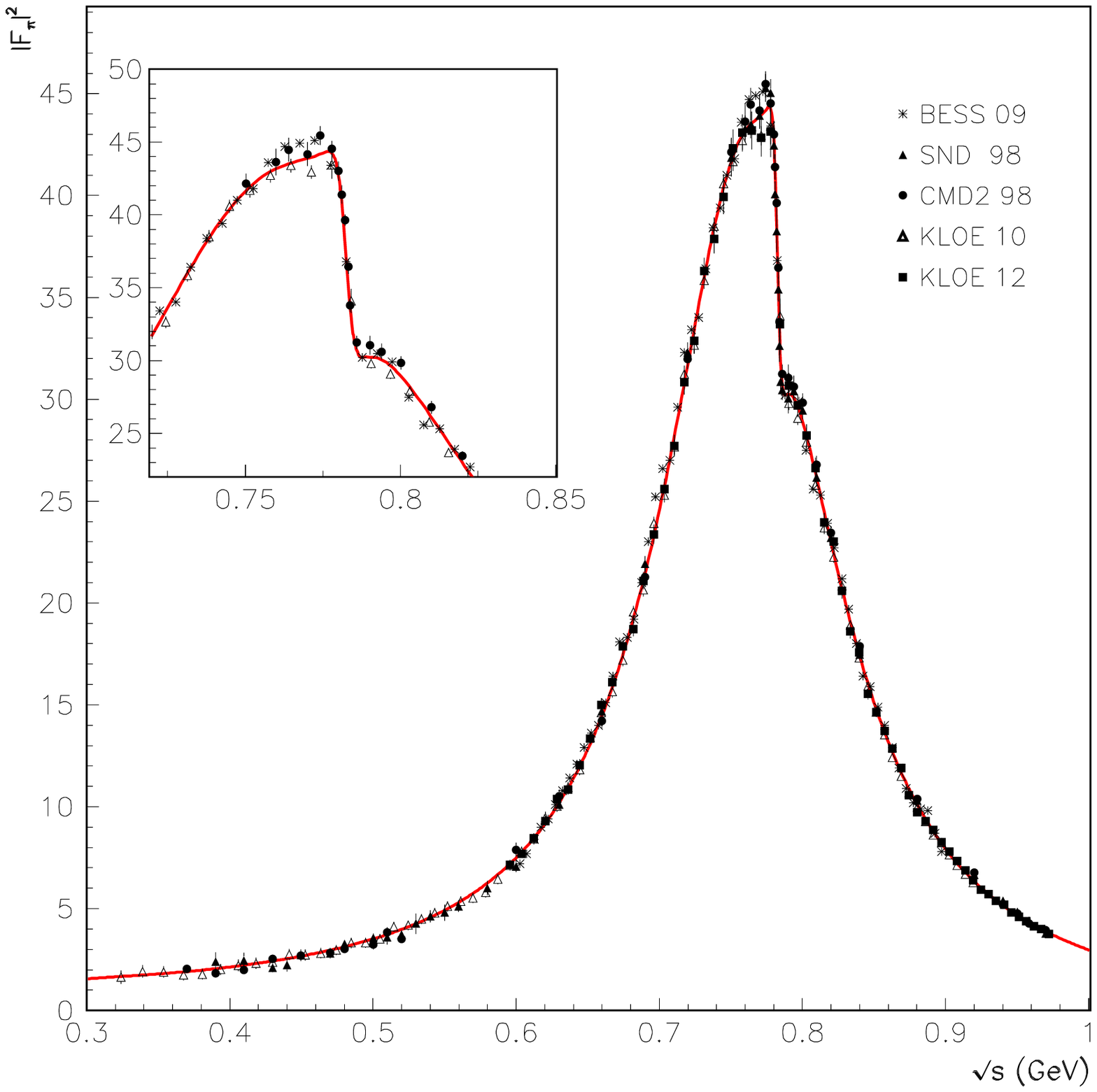}
\hspace{2pc}%
\includegraphics[width=8cm]{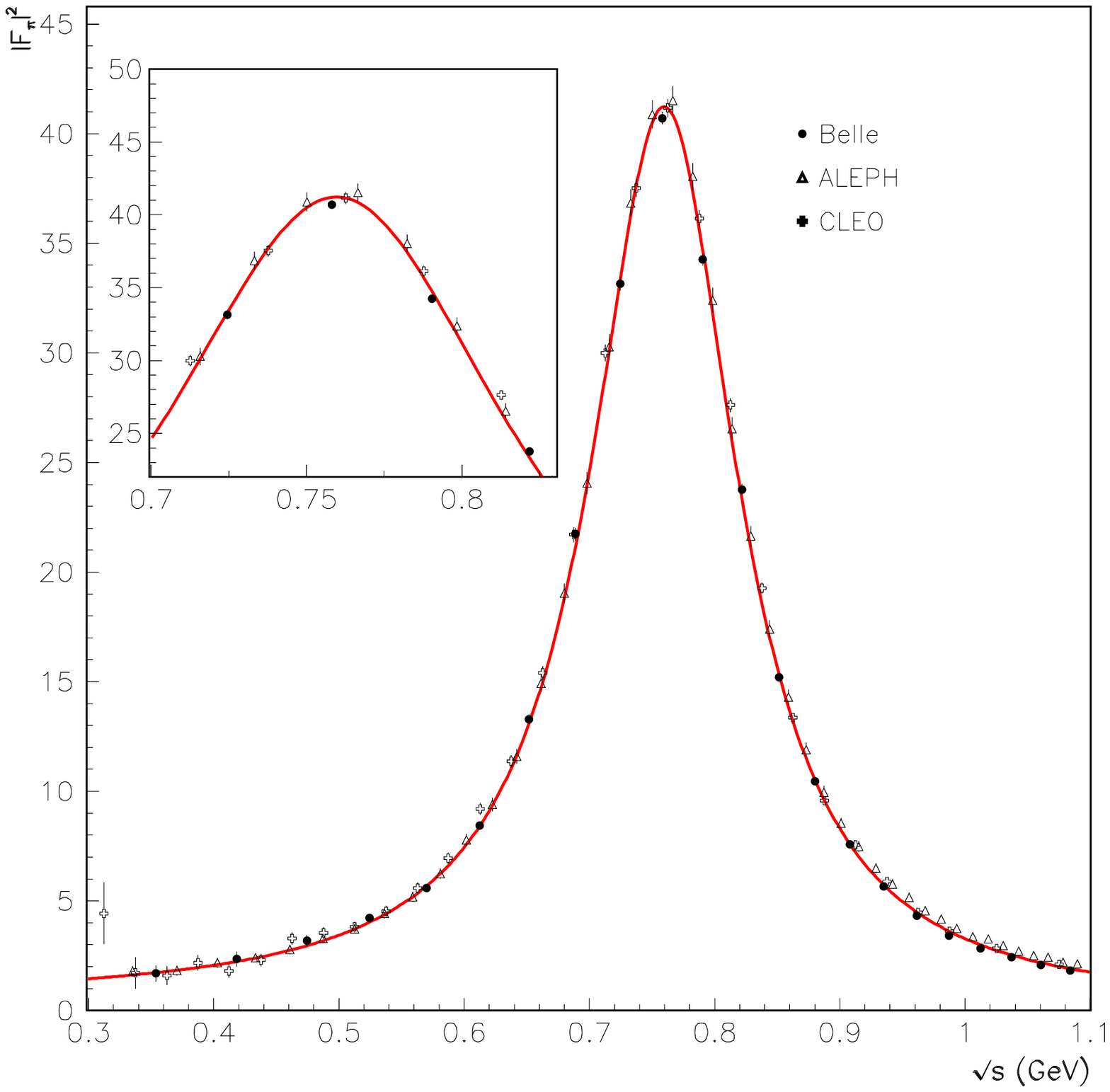}
\caption{ The pion form factor in the  $e^+e^- \to \pi^+\pi^-$  
 annihilation (left--hand pannel) and in the $\tau$ decay (right--hand pannel) 
 derived from a BHLS fit involving the CMD--2, SND, KLOE10, KLOE12 and BESS  data samples,
 on the one hand, and  the ALEPH, CLEO and Belle dipion spectra, on the other hand.  }
\label{tau_ee_fit}    
\end{figure*}

The first data line in Table \ref{Table:T1} reports some fit information 
derived using the various $\pi^+\pi^-$ samples in isolation, namely their various
$\overline{\chi}^2$ and their global fit probabilities\footnote{As illustrated
by Table 3 in \cite{ExtMod3} and reminded in \cite{ExtMod4,ExtMod5}, the global fit
probabilities are enhanced towards 1 because several groups of data samples --
especially those collected in the  $\pi^0 \g$ and $\eta \g$ channels -- benefit
from very favorable partial $\chi^2$. Under these conditions, small global
 probabilities indicate suspicious behaviors.}; in these fits, the PDG information
 previously referred to should be removed.  One observes a significant gap between
 KLOE08 and BaBar, on the one hand
   and the five other data samples, on the other 
 hand\footnote{The quantity denoted $\chi^2_{\pi^+\pi^-}$ is, of course, the 
 contribution to the total $\chi^2$ of the $\pi^+\pi^-$ data.}.

 In the same Table, one also displays the fit results
associated with different combinations of the existing data samples. This
Table shows that the
largest set where each data sample has an average $\chi^2$ per point close to
its value in its single mode fit is Combination 1; this combination is our reference
for the following. For this combination,
the $\overline{\chi}^2$ of the whole set of $\pi^+\pi^-$ data sample is 0.98
with an associated large probability as indicated in the first column.

The pion form factor (FF) in $e^+e^-$ annihilations and in the $\tau$ decay derived from fitting
with Combination 1 are displayed in Figure \ref{tau_ee_fit}; they  are clearly satisfactory. 
The older $\pi^+\pi^-$ data reported in \cite{Barkov} are also included in the fit and, on the whole,
one yields   $\chi^2_{\pi^+\pi^-}/N_{\pi^+\pi^-}= 361.5/404$ 
and $\chi^2_{\pi^\pm\pi^0}/N_{\pi^\pm\pi^0}= 86.8/85$.
 Therefore the picture looks satisfactory and this should be reflected
 by the residual plots.

\begin{figure}
\centering
\includegraphics[width=8cm,clip]{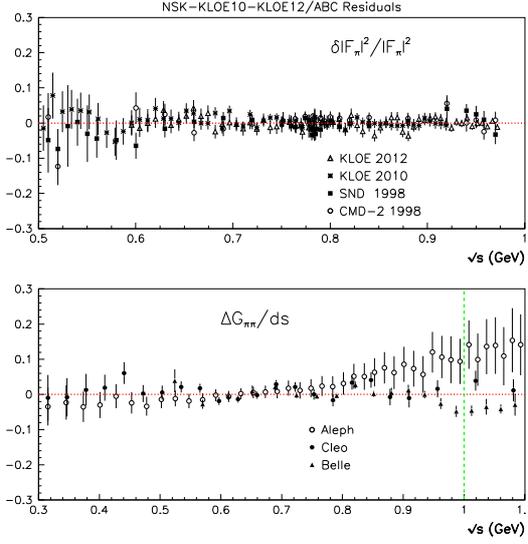}
\caption{ Residuals for the  pion form factor in the  
$e^+e^- \to \pi^+\pi^-$   annihilation (upper pannel) and in the $\tau$ decay 
(lower pannel). The residuals in the upper pannel are corrected for the global scale uncertainty
effects (see \cite{ExtMod4,ExtMod5}).}
\label{tau_ee_res}    
\end{figure}

   Figure \ref{tau_ee_res} displays  the  $e^+e^-$ pion FF residuals 
appropriately corrected for the reported scale uncertainty 
effects as discussed in the Appendix of \cite{ExtMod4}
and in \cite{ExtMod5}; it looks reasonably flat. 

 The pion FF in  the $\tau$ decay is flat
for the CLEO and Belle spectra\footnote{Actually, the 3 residual distributions
 shown in the lower pannel of Figure \ref{tau_ee_res}  look quite similar to the Belle 
 plots displayed in Figure 12 of \cite{Belle}.}; 
 the ALEPH data sample tends to exhibit a  small growth
starting at $\simeq 850$ MeV. However, Figure 3 in \cite{AlephCorr} indicates that this 
distribution, thanks to a bug fix, 
should be scaled down just in this energy region  and, then, it behaves like the others.

Therefore, the global fits mixing the $\tau$ spectra and the  $e^+e^- \to \pi^+\pi^-$  
data samples confirm the conclusions reached in the previous Section with the $\tau$ + PDG
method. Stated otherwise, global fits do not indicate any mismatch between
the $e^+e^-$ and $\tau$ spectra within BHLS.

\section{Including \& excluding the $\tau$ spectra~: Hadronic HVP issues}
\label{hvp}
\begin{figure} 
\centering
\includegraphics[width=8cm,clip]{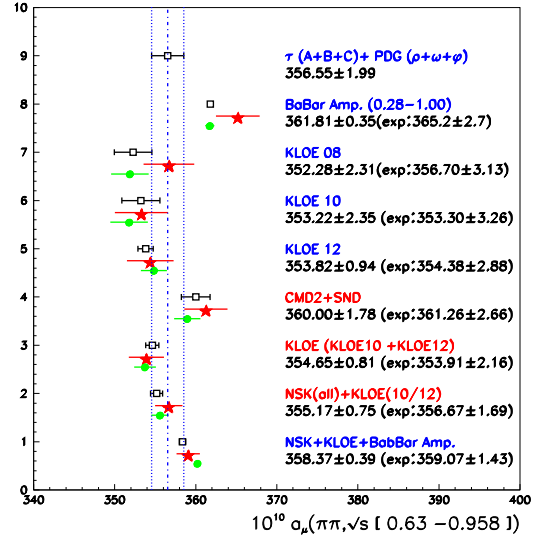}
\caption{ Values for $a_\mu(\pi \pi,[0.63,0.958])$ in units
of 10$^{-10}$ derived from global fits using the indicated $e^+e^-\ra \pi^+ \pi^-$ 
data samples or combinations; the $\tau$ dipion spectra are always used.
The full green circles are the results obtained from the $A=m$ fit (no iteration) and 
the  black empty squares are the results obtained from the $A=M_0$
fit (first iteration) s explained  in \cite{ExtMod5}.
The values derived by directly integrating the experimental spectra are 
 indicated by red stars. See Section \ref{hvp} for comments.}
\label{Fig:a_mu_ref}
\end{figure}

\indent \indent
Let us first examine the contributions to the muon HVP provided by the pion loop
in the energy region $[0.630,0.958]$ GeV; Figure \ref{Fig:a_mu_ref} displays our results.
The point at top of this Figure displays the $\tau$+PDG prediction for 
 $a_\mu(\pi \pi,[0.63,0.958])$.
The data points in red display the corresponding information directly reconstructed from
the samples provided by the indicated experiments; some combinations of these are also
shown. So, one observes a quite good correspondance between the "experimental" values
and the prediction derived by the $\tau$+PDG method; of course, the experimental values
are not influenced at all by BHLS or the $\tau$ data.

In the same Figure, one also displays the results derived when merging
 the  $e^+e^- \to \pi^+\pi^-$ data samples and the $\tau$ dipion spectra in 
the minimization procedure. The empty black symbols show the fit results
derived by using the iterative method defined in \cite{ExtMod5}. 
The points in green are the corresponding results derived from the same
fits but performed without iterating. The motivation for an iterative method
are emphasized in \cite{ExtMod5} and  aims at cancelling out possible biases 
affecting the channels dominated
by samples subject to dominant global scale uncertainties

 Figure \ref{Fig:a_mu_ref} shows that the $\tau$+PDG prediction as well as the fits
 merging $\tau$ and $e^+e^-$ data are consistent with each others and also with
 the experimental data, except for BaBar which has difficulties to accomodate the BHLS
 framework as shown in Table \ref{Table:T1}.  Because of the energy boundaries of its
 spectrum, a BESSIII experimental datum for  $a_\mu(\pi \pi,[0.63,0.958])$ cannot be
 produced.

Let us go a step further and examine the contributions to the muon HVP
accessible through the BHLS Lagrangian and fitter which is, as already stated, limited upward
slightly above the $\phi$ mass; we chose 1.05 GeV. The results are displayed 
in Table \ref{Table:T2}. The numbers have been derived using the iterated fit method
already referred to \cite{ExtMod5}.

Let us focus on the $\pi^+ \pi^-$ contribution which is actually the main aim
of the present study. One thus observes that including the $\tau$ spectra shifts
the central value by about $1.5 \times 10^{-10}$ and improves the uncertainty by
about $0.3 \times 10^{-10}$ in both cases.
The $1.5 \times 10^{-10}$ difference between excluding and including the $\tau$ spectra
 looks rather small ($\simeq 1~\sigma$ or less), similar to those obtained in 
\cite{Fred11}, but much smaller than
 those in  \cite{DavierHoecker3}. Therefore, within BHLS, the contribution  of the 
 $\pi^+ \pi^-$ channel to the HVP does not exhibit any singular behavior~: Using
 or not the $\tau$ spectra  does not change this picture but improves the results as 
 expected from having a larger statistics  (e.g. the $\pi^+ \pi^-$ $and$ the $\tau$ data).

\begin{table*}
\caption{ The various contributions to the muon HVP $a_\mu$ in units of $10^{-10}$
using the BHLS fit excluding the $\tau$ data (first data column), including the $\tau$ 
spectra (second data column), compared with the direct integration of the experimental
 spectra (last data column).  The last line displays the
 total contribution accessible through the BHLS Model. The $e^+e^-$ data samples involved
 are those from CMD--2, SND, KLOE10, KLOE12 and BESSIII.
} 
\centering
\label{Table:T2} 
\hspace{-0.7cm}
\begin{tabular}{llll}
\hline                              
Channel & ~~~~Excl. $\tau$ & ~~~~Incl. $\tau$  & Direct Estim. \\ 
\hline                              
$\pi^+ \pi^-$   &$493.02 \pm 1.16$  &$494.59 \pm 0.89$  &$492.98 \pm 3.38$ \cr
\hline                              
$\pi^0 \g$ & ~~~$4.50\pm 0.05$ &  ~~~~$4.54\pm 0.04$ &  ~~~$3.67\pm 0.11$    \\
$\eta \g$ & ~~~$0.64\pm 0.01$ &  ~~~$0.64\pm 0.01$ &  ~~~$0.56\pm 0.02 $   \\
$\pi^+ \pi^-\pi^0$ & ~$40.84\pm 0.62$ &  ~$40.84\pm 0.57$ &  ~$43.54\pm 1.29$ \\ 
$K_L K_S$ & ~$11.53\pm 0.09$ &  ~$11.53\pm 0.08$ &  ~$12.21\pm 0.33$  \\   
$K^+ K^-$  & ~$16.88\pm 0.22$ &  ~$16.90\pm 0.20$ &  ~$17.72\pm 0.52$  \\[0.2cm]   
Total ($<1.05$ GeV)  &$567.00 \pm 1.63$  &$569.04 \pm 1.08$  &$570.68\pm 3.67$ \\ 
\hline                              
\end{tabular}
\end{table*}

\section{Conclusion}
\label{conclusion}
\indent \indent  The analysis developped  above  leads to conclude
that, actually, one does not observe any mismatch between the $e^+e^-$ and the $\tau$ data. To be as precise 
as possible, Table \ref{Table:T1} indicates that the $\tau$ spectra collected by ALEPH, CLEO 
and Belle are in perfect agreement with the CMD--2, SND, KLOE10, KLOE12 and BESSIII data 
samples\footnote{e.g. five out of the seven high statistics existing data samples.
}
each taken in isolation or considered together within a sample combination. 
This leads us to conclude that the so--called $\tau - e^+e^-$ puzzle is only due to 
the way the implementation of   isospin symmetry breaking is performed within some
 models. In contrast, the BHLS approach and its way to account for IB effects
seem to reflect correctly the expected relationship
and the expected closeness of the  $e^+e^-$ annihilation and $\tau$ decay processes. Indeed,
BHLS provides successful $\tau$ predictions of the $e^+e^-$ pion form factor and a
good simultaneous fit of both kinds of data.

However,
we are left with a significant tension between the KLOE08 and BaBar (up to 1 GeV) samples 
on the one hand and the $\tau$ spectra on the other hand, as well reflected by the $\tau$+PDG
information collected in Figure \ref{tau_pred} and by the BHLS  global fit results displayed
in Table \ref{Table:T1}. This is indeed an issue but, seemingly, external to the so--called $\tau - e^+e^-$ puzzle 
which motivates this work.

%


\begin{thebibliography}{}

\bibitem{puzzle1}  Davier M, Eidelman S, Hoecker A and Zhang Z
               2003 {\it Eur.Phys.J.} C {\bf 27} 497

\bibitem{puzzle2}  Davier M, Eidelman S, Hoecker A and Zhang Z
               2003 {\it Eur.Phys.J.} C {\bf 31} 503

\bibitem{Davier2007}	Davier M 2007 	{\it Nucl. Phys. Proc. Suppl. }	 {\bf 169} 288

\bibitem{DavierHoecker3} Davier M, Hoecker A, Malaescu B  and Zhang Z  2011 
    			 {\it Eur.Phys.J. } C {\bf 71} 1515 (preprint arxiv:1010.4180)
\bibitem{taupaper} 
     Benayoun  M, David P, DelBuono L, Leitner  O  and O'Connell H B  2008
	{\it  Eur.Phys.J. } C {\bf 55} 199-236 (preprint hep-ph/0711.4482)

\bibitem{Fred11} Jegerlehner F and Szafron 2011 
{\it Eur. Phys. J.} C {\bf 71} 1632 (preprint arXiv:1101.2872)


 \bibitem{DavierHoecker} Davier M, Hoecker A,  Lopez Castro G, Malaescu B, Mo, X H,
 Toledo Sanchez G, Wang P,Yuan C Z  and Zhang Z
         2010 {\it Eur. Phys. J.} C
           {\bf 66} 127--136 (preprint arxiv:0906.5443) 
	   
\bibitem{HLSRef}  Harada M and Yamawaki K  2003 
	{\it Phys. Rept.} {\bf 381} 1-233 (preprint hep-ph/0302103)

\bibitem{ExtMod3} Benayoun M, David P, DelBuono, L and Jegerlehner F  2012
       {\it Eur.Phys.J.} C {\bf 72} 1848 (preprint arXiv:1106.1315)

\bibitem{FKTUY} Fujiwara T, Kugo T, Terao, H, Uehara S and Yamawaki K  1985
{\it Prog. Theor. Phys.} {\bf 73}  926--941

\bibitem{BKY}  Bando  M, Kugo  T and Yamawaki, K 1985
	 {\it Nucl. Phys.} B {\bf 259} 493--502

\bibitem{Hashimoto} Hashimoto  M 1996	 {\it Phys. Rev.} D {\bf 54} 5611-5619

\bibitem{Heath} Benayoun M  and O'Connell H  B 1998 {\it Phys. Rev.} D 074006 

\bibitem{tHooft} 't Hooft G 1986 {\it Phys. Rept.} {\bf 142 } 357-387

\bibitem{Marciano} Marciano W J and Sirlin A 1993  {\it Phys. Rev. Lett.}
     {\bf71} 3629-3632

\bibitem{Cirigliano} Cirigliano V, Ecker  G and Neufeld H 2001
                      {\it Phys. Lett} B {\bf 513} 361-370 (preprint hep-ph/0104267)
\\  Cirigliano V, Ecker  G and Neufeld H 2002
		{\it JHEP} {\bf  08}  002  (preprint hep-ph/027310)

\bibitem{ExtMod4} Benayoun M, David P, DelBuono, L and Jegerlehner F  2013
       {\it Eur.Phys.J.} C {\bf 73} 2453 (preprint arXiv:1210.7184)

\bibitem{RPP2012} Beringer J {\it et al.} 2012   Review of Particle Physics (RPP)
  		{\it Phys.Rev.} D {\bf 86} 010001

\bibitem{CMD-2} Aulchenko V  M {\it et al.} 2002  
		{\it Phys. Lett.} B {\bf 527} 161-172  (preprint hep-ex/0112031)
\\ Akhmetshin R  R {\it et al.}  2006 
		{\it Phys. Lett.} B {\bf 648} 28-38  (preprint hep-ex/0610021)
\\  Akhmetshin R  R {\it et al.}  2006 
		{\it JETP Lett.}   {\bf 84} 413-417  (preprint hep-ex/0610016)
\bibitem{SND} Achasov  M N {\it et al.} 2006 
		{\it JJ. Exp. Theor. Phys.}   {\bf 103} 380-384  (preprint hep-ex/0605013)


\bibitem{KLOE08} Venanzoni  G {\it et al.}  2009 {\it AIP Conf. Proc.} {\bf 1182} 665
			(preprint arXiv:0906.4331)

\bibitem{KLOE10} Ambrosino F {\it et al.} 2011 {\it Phys. Lett.} B {\bf 700} 102-110
			(preprint arXiv:1006.5313)

\bibitem{KLOE12} Babusci D {\it et al.}	2013  {\it Phys. Lett.} B {\bf 720} 336-343
			(preprint arXiv:1212.4524)
			
       
 \bibitem{BaBar}   Aubert  B {\it et al.} 2009	{\it Phys. Rev. Lett. }
        {\bf 103} 231801 (preprint arXiv:0908.3589)
\\  Lees  J P {\it et al.} 2012 {\it Phys. Rev.  } D {\bf 86} 032013
           (preprint arXiv:1205.2228)
\bibitem{BESS-III}   Ablikim M  {\it et al.} 2015 (preprint arXiv:1507.08188)
       
\bibitem{Conf_2013} Benayoun M  2014 {\it Int.J.Mod.Phys.Conf.Ser.}
			{\bf 35} 1460416
 \\ Benayoun M  2013 {\it PoS} {\bf Photon2013} 048
 
 \bibitem{ExtMod5} Benayoun M, David P, DelBuono, L and Jegerlehner F  2015
       (preprint arXiv:1507.02943)

\bibitem{Aleph}  Schael S {\it et al.}  2005  {\it Phys. Rept. } 
        {\bf 421} 191-284 
\bibitem{Cleo}   Anderson S {\it et al.}  2000 {\it Phys. Rev.} D {\bf 61} 112002
				(preprint hep-ex/9910046)
\bibitem{Belle}  Fujikawa M {\it et al.} 2008 {\it Phys. Rev.} D {\bf 78} 072006
				(preprint arXiv:0805.3773)	
	   
\bibitem{Barkov} Barkov L M  {\it et al.} 1985 {\it Nucl. Phys.} B  {\bf 256} 365-384

\bibitem{AlephCorr}  Davier  M,  Hoecker  A,  Malaescu  B,  Yuan C--Z and Zhang Z  2014
			 {\it Eur.Phys.J. } C {\bf 74} 2803 (preprint arXiv:1312.1501)



\end{thebibliography}
\end{document}